\documentclass[a4paper]{jpconf}
\usepackage{citesort}
\usepackage{amsmath}
\usepackage{amsfonts}
\usepackage{amssymb}
\usepackage{graphicx}

\begin{document}

\title{Dynamic Formation of Metastable Intermediate State Patterns in Type-I Superconductors}

\author{R Prozorov and J R Hoberg}

\address{Ames Laboratory and Department of Physics \& Astronomy, Iowa State
University, Ames, IA 50011, U.S.A.}

\ead{prozorov@ameslab.gov}

\begin{abstract}
Structure of the intermediate state in type-I superconducting lead (Pb) is shown to be very sensitive to the ramp rate of an applied magnetic field. The configurations of resulting static patterns depend sensitively on the shape of the specimen. In particular, geometric barrier, present in the samples with rectangular cross-section, plays an important role in determining the sharp boundary between the phases of different topology. We propose that seemingly laminar (stripe) pattern obtained as a result of the fast field ramp is simply an imprint left behind by the fast-moving flux tubes. Our results confirm that flux tube phase is topologically favorable.
\newline\newline
\textit{Date: 14 June 2008}
\end{abstract}

Topology of the intermediate state (IS) in type-I superconductors is an important subject that deals with both, the fundamental physical properties and behavior of complex strongly correlated systems \cite{Huebener2001,parks,Prozorov2007,Prozorov2008}. Macroscopically, a sample must obey well-established thermodynamic and electrodynamic laws of type-I superconductivity. On the other hand, pattern topology is not determined solely by the energy minimization, but depends sensitively on many parameters, including sample purity, pinning, shape and magnetic protocol used to induce the IS. This leads to various interesting phenomena such as topological hysteresis\cite{Prozorov2007}. On flux penetration into the Meissner state, tubes carrying many flux quanta snap off the edges and move towards the center, gradually forming a well-defined honeycomb structure called "suprafroth". The suprafroth coarsens and obeys general statistical laws just as conventional soap froth\cite{Prozorov2008}. However, the question is whether this topology is robust with respect to the motion caused by the Lorentz force due to shielding currents.

\begin{figure}[h]
\begin{center}
\includegraphics[width=15cm]{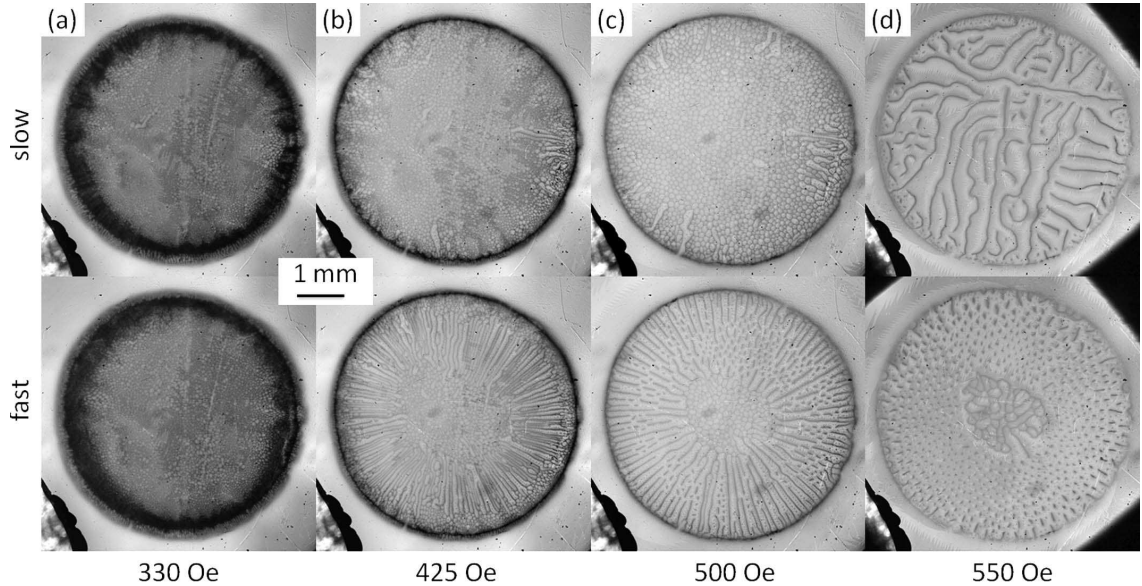}
\end{center}
\caption{\label{fig1}Intermediate state patterns in a disc-shaped lead single crystal upon flux penetration at $T=5$ K. Top row shows IS obtained with a slow ramp ($\sim 10$ Oe/s) at the indicated magnetic fields. Bottom row shows fast ramp ($\sim 3000$ Oe/s) structures imaged at the same field. Note that all images are static and do not evolve in time as long as $H$ and $T$ are kept constant.}
\end{figure}

Dynamics of the IS was studied in a number of previous works, including direct magneto-optical visualization \cite{Huebener2001,parks,DeSorbo1965,Haenssler1965,Solomon1971,Dutoit1989,Castro1999}. Electric current - driven IS evolving into a regular stripe pattern was demonstrated \cite{Huebener2001,parks,DeSorbo1965,Haenssler1965,Solomon1971,Dutoit1989}. Related experiments explored fast switching magnetic field and they were understood in terms of fast moving flux tubes that create these ordered patterns. In some cases, Magnus force acting on the tubes even caused "flux precession"\cite{parks,DeSorbo1965,Haenssler1965}. Recent work, surprisingly, did not offer any physical interpretation of similar results obtained in In films (and only marginally regular pattern was actually shown) \cite{Jeudy2006}.

In this paper we examine the effect of fast magnetic field ramp in samples of different shapes and show that IS pattern depends sensitively on the ramp rate forming metastable stripe phase different from both the suprafroth and the Landau laminar structure. This dynamic pattern remains stable after the field ramp, indicating that the energy difference between different topologies is negligible and that topological features are determined by the flux channeling and mobility rather than free energy minimum.

In the experiment, a component of the magnetic induction perpendicular to the surface of the sample was measured by using magneto-optical imaging technique that utilizes rotation of light's polarization plane (Faraday effect) in bismuth-doped iron-garnet indicators with in-plane magnetization \cite{Prozorov2007}. A flow-type liquid $^4$He cryostat with sample in vacuum was used. Sample was fixed on top of a copper cold finger with the small amount of Apiezon grease and an indicator was placed on top of the sample. When a linearly polarized light passes through the indicator and reflects off the mirror sputtered on its bottom, it picks a double Faraday rotation proportional to the magnetic field intensity at a given location on the sample surface. Observed through the (almost) crossed analyzer, we recover a 2D image where intensity is proportional to the local magnetic induction. Magnetic field was applied with a conventional solenoid with a computer-controlled slowest ramp rate of $\sim 10$ Oe/s and the highest of $\sim 3000$ Oe/s.

\begin{figure}[h]
\includegraphics[width=9cm]{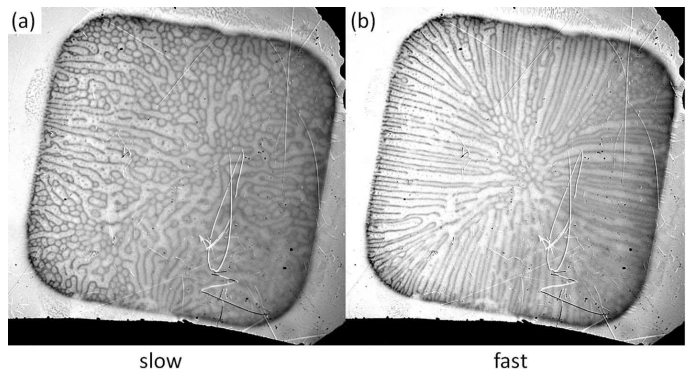}\hspace{2pc}%
\begin{minipage}[b]{5.5cm}\caption{\label{fig2}Fast penetration into a square shaped slab (right) compared to the slow penetration (left). Note remanent flux tubes along the diagonals - places with smallest driving Lorentz force. $T=5$ K.}
\end{minipage}
\end{figure}

Figure \ref{fig1} shows patterns obtained in a disc-shaped single crystal after slow (top row) and fast (bottom row) magnetic field ramps. While the slow pattern shows regular flux tube structure, the fast ramp reveals radial lines. Physically, when magnetic field is increased, shielding current increases as well, so there is an increasing driving Lorentz force on the flux tubes. An interesting point, however is that Fig.~\ref{fig1} (bottom row) shows that there is a sharp boundary between the fast propagating tubes that create flux channels and the regular tubular structure. What happens is that first tubes move very fast into the sample center and their density grows towards the edges as the field increases. At some point they collide with the incoming tubes and this defines the boundary between the two phases and allows to estimate the velocity of fast moving tubes of about $10^3$ m/sec. Another very interesting observation is the last image of the bottom row of Fig.~\ref{fig1} where fast ramp has apparently produced a lattice of superconducting tubes. (The flux tubes we discussed above consist of normal regions in the superconducting environment. Fig.~\ref{fig1} shows the opposite - superconducting tubes ordered in a lattice in the normal phase).

\begin{figure}[h]
\begin{center}
\includegraphics[width=13cm]{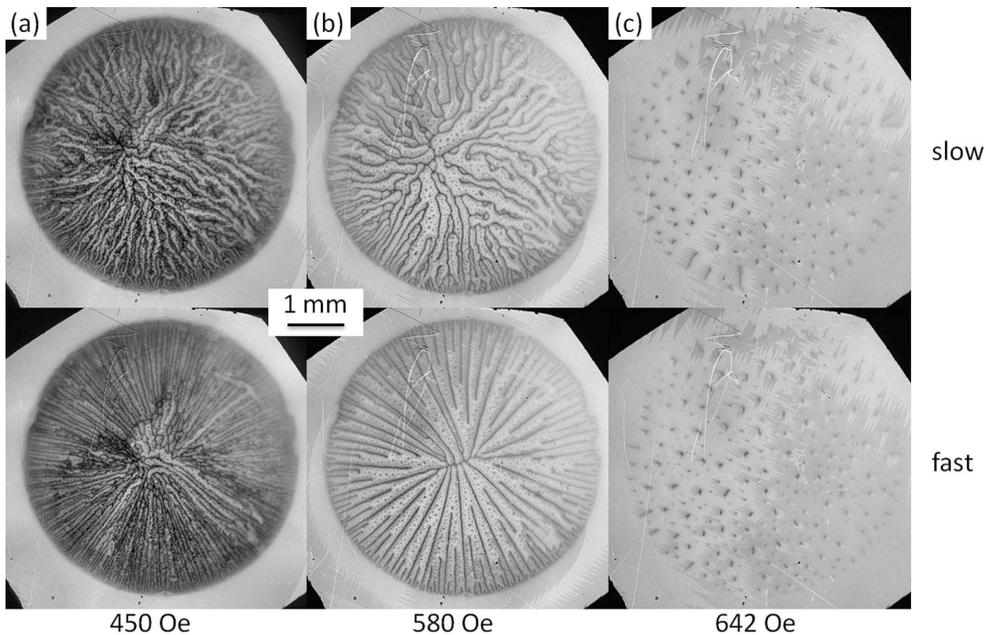}
\end{center}
\caption{\label{fig3}Intermediate state in a hemisphere. Top row is the slow ramp, bottom row is the fast ramp.}
\end{figure}

\begin{figure}[h]
\begin{center}
\includegraphics[width=10cm]{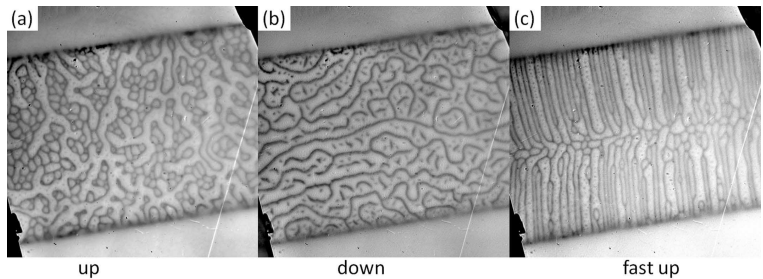}
\end{center}
\caption{\label{fig4}IS in the long strip. (a) shows a slow ramp up after zfc, (b) is the ramp down after going to the normal state and (c) is the fast ramp up of the magnetic field after zfc.}
\end{figure}

We now examine how sample shape influences the resulting topology of the fast penetrating flux. Figure \ref{fig2} shows fast field ramp (right) penetration compared to the slow ramp (left) in a rectangular slab. The pattern follows the Lorentz force, which balances between external square geometry and internal (far from the boundaries) cylindrical symmetry point. When the cross-section is changed and the geometric barrier is removed, the situation changes. Figure \ref{fig3} shows the intermediate state in a hemisphere. Top row is the slow ramp, bottom row is the fast ramp. There is a clear difference between a hemisphere and the disc, Fig.~\ref{fig1}. No "tube" zone is formed, which is simply because tubes do not penetrate inside and the Lorentz driving force is compensated by the superconducting condensation energy increase when the tube length increases \cite{Prozorov2007}. Finally, we show fast penetration of the magnetic flux into a long strip. Figure \ref{fig4} confirms previous findings that the IS pattern formed after fast ramp consists of parallel superconducting walls with their direction perpendicular to the sample edge.

The physical interpretation of the observed phenomena is a follows. When a flux tube is nucleated at the sample edge, it is moved in by the Lorentz force due to generated Meissner current. When a magnetic field is ramped fast, the Meissner current density increases rapidly exerting an increasing force on the flux tube, which, in turn, causes fast acceleration and ultimately reaching terminal velocity determined by the viscosity. So, the "rivers of flux tubes" (or "trains of tubes" as John Clem called them) are moving very fast perpendicular to the current flow (as required by the cross product in the Lorentz force expression, $\bf{F_L}=\bf{v}\times\bf{\Phi}$ where $\bf{v}$ is tube velocity and $\bf{\Phi}$ is the flux curried by the tube). In the case of a a disc, tubes move even faster, because external ramp rate is slower that un-inhibited by the condensation energy loos accelerated motion. In the case of a hemisphere, however, they move against the growing counter-force of the condensation energy due to increasing tube length.

An important conclusion is that that fast field ramps show that tubular structure is, indeed, the equilibrium intermediate state in type-I superconductors. Tubes have much better topological mobility compared to the Landau laminar structure. Pinning and imperfections break the tubular structure, but flux motion lifts the tubes off their pinning wells and moving IS consists of tubes that manifest themselves as regular traces left behind.

\ack
Work at the Ames Laboratory was supported by the Department of Energy-Basic Energy Sciences under
Contract No. DE-AC02-07CH11358. R. P. acknowledges support from NSF grant number DMR-05-53285 and the Alfred P. Sloan Foundation.

\end{document}